\documentclass
[prl,showpacs,superscriptaddress,preprint,prl,12pt,onecolumn]{revtex4}%
\usepackage{amsfonts}
\usepackage{amsmath}
\usepackage{amssymb}
\usepackage[dvips]{graphicx}%
\begin{document}
\author{Gerrit E. W. Bauer}
\affiliation{Kavli Institute of NanoScience, Delft University of Technology, Lorentzweg 1,
2628 CJ Delft, The Netherlands}
\author{Yaroslav Tserkovnyak}
\affiliation{Lyman Laboratory of Physics, Harvard University, Cambridge, Massachusetts
02138, USA}
\author{Arne Brataas}
\affiliation{Department of Physics, Norwegian University of Science and Technology, N-7491
Trondheim, Norway}
\author{Jun Ren}
\affiliation{Chinese Academy of Sciences, Beijing, China}
\author{Ke Xia}
\affiliation{Chinese Academy of Sciences, Beijing, China}
\author{Maciej Zwierzycki}
\affiliation{Faculty of Science and Technology and MESA+ Research Institute, University of
Twente, 7500 AE Enschede, The Netherlands}
\author{Paul J. Kelly}
\affiliation{Faculty of Science and Technology and MESA+ Research Institute, University of
Twente, 7500 AE Enschede, The Netherlands}
\altaffiliation[Permanent address: ]{Institute of Molecular Physics, P.A.S., Smoluchowskiego 17, 60-179 Pozna\'n, Poland}

\title{Spin accumulation and decay in magnetic Schottky barriers}

\begin{abstract}
The theory of charge and spin transport in forward-biased Schottky barriers
reveals characteristic and experimentally relevant features. The conductance
mismatch is found to enhance the current induced spin-imbalance in the
semiconductor. The GaAs%
$\vert$%
MnAs interface resistance is obtained from an analysis of the magnetic field
dependent Kerr rotation experiments by Stephens \textit{et al.} and compared
with first-principles calculations for intrinsic interfaces. With increasing
current bias, the interface transparency grows towards the theoretical values,
reflecting increasingly efficient Schottky barrier screening.

\end{abstract}

\date{\today}
\maketitle

An obstacle to the direct injection of spins from a ferromagnetic metal (F)
into a semiconductor (SC) is the so-called conductance mismatch
\cite{Schmidt00}. Paradoxically, this problem is most severe for good electric
contact, because most of the applied potential drop is then wasted over the
highly resistive semiconductor and very little is left to spin-polarize the
current in the magnetically active region. Selection rules at ballistic
interfaces are responsible for a large interface spin polarization
\cite{Wunnicke, Zwierzycki} that allows significant spin accumulation in spite
of the mismatch, but even small amounts of disorder have detrimental effects
\cite{Zwierzycki}. Use of ferromagnets with low conductances matched to those
of the semiconductor \cite{Fiederling99} is another route, but many material
problems, such as low critical temperatures for ferromagnetism, have still to
be solved \cite{Ohno}. Spins can be effectively injected into a semiconducting
base contact of a three-terminal spin-flip transistor \cite{Brataas00} or by
pumping spins into the semiconductor by ferromagnetic resonance
\cite{Brataasb}, but none of these theoretical predictions have been confirmed
experimentally yet. The spin polarization of the injected current can be
increased by tunneling or Schottky barriers, both causing the applied
potential to drop in the spin-selective region of the sample \cite{Schmidt00,
Rashba00, Fert01}. This feature has been employed in experiments that divide
into two categories. In the first, hot electrons are injected into a metallic
magnetic multilayer base in the forward bias regime. In the \textquotedblleft
spin-valve transistor\textquotedblright\ \cite{Monsma} this is achieved via a
Schottky barrier; in \textquotedblleft magnetic tunnel
transistors\textquotedblright, tunneling barriers are used instead
\cite{Parkin}. The second category of experiments concentrates on injecting
spins from the ferromagnet into semiconductors by applying a reverse bias,
reaching polarizations of 30
$\backslash$%
\% \cite{Jonker}. Here the \emph{spin current} is the observable, measured by
the circular polarization of the recombination luminescence of the injected
electrons with thermalized holes.

Recently, Stephens \textit{et al}. \cite{Stephens} investigated in
forward-biased Schottky barriers not the hot electrons that traverse a
ferromagnetic base as in Refs. \cite{Monsma,Parkin}, but the cold ones that
remain in the semiconductor. A significant bias-dependent \emph{spin
accumulation} in the semiconductor was observed by Kerr rotation. The
interpretation as spin dependent reflection at the interface was supported by
a simple parabolic band/step potential model. In this Letter we present a
theoretical analysis based upon an adaptation of magnetoelectronic circuit
theory \cite{Brataas00,Bauer04}. We find that the conductance mismatch has a
beneficial effect on the size of the spin accumulation. Analyzing the Bloch
equation that governs the spin accumulation in the presence of an applied
magnetic field we find that the experimental results on the dephasing by a
magnetic field (Hanle effect) can be used to extract the SC%
$\vert$%
F interface resistance. We also present first-principle calculations of
intrinsic interface transport parameters for disordered interfaces as a
function of the SC Fermi energy.

The sample configuration is indicated in Fig. 1. We start with a discussion of
an infinite planar Schottky barrier model between a degenerately n-doped
semiconductor SC and a metallic ferromagnet F that is kept at low temperatures
and biased with an electric particle current $I_{C}$. With increasing forward
(positive) bias the semiconductor band edge is lifted relative to the
ferromagnetic one. The ionized donor atoms are increasingly screened until at
a bias close to the Schottky barrier height the semiconductor band edge at the
interface comes close to the bulk Fermi energy of the semiconductor
$\varepsilon_{F}$. The \textquotedblleft flat-band\textquotedblright%
\ condition is defined asymptotically at a voltage close to the barrier height
where the tunneling current and thus the electric field in the semiconductor
start to become significant. Although the theory is valid for arbitrary
material combinations we concentrate here on the sample investigated by
Stephens \textit{et al.}, in which the GaAs is n-doped with densities of
$\sim10^{17}\mathrm{%
\operatorname{cm}%
}^{-3}$.$\ $With an impurity scattering mean free path of $\sim30$ $%
\operatorname{nm}%
$ the semiconductor is safely in the diffuse transport regime. The I-V
characteristic in the forward bias shows a band tail close to the Schottky
barrier and roughly Ohmic behavior at high bias with a resistance of 300 $%
\operatorname{\Omega }%
$, indicating that the thin semiconductor layer limits the transport.
Important parameters are the spin-flip diffusion length of $\ell_{\mathrm{sd}%
}\simeq2$ $%
\operatorname{\mu m}%
$ \cite{Kikkawa}, and a flat-band depletion length of $\sim20$ $%
\operatorname{nm}%
$. Any residual band-bending is thus incorporated in the (quantum) interface
resistance. The conductance of the high-density metallic ferromagnet MnAs is
much higher than that of the semiconductor and disregarded. We concentrate on
the dimensionless spin-dependent $\left(  s=\uparrow,\downarrow\right)  $
occupation function $f_{s}\left(  \varepsilon\right)  $ in the semiconductor
near the interface at an energy $\varepsilon$ from the band edge. The up spin
direction $\uparrow$ is chosen parallel to the majority spin in the
ferromagnet. Close to the flat band condition the energy of the electrons
entering the metal is of the order of the Schottky barrier height, that is
much larger than the semiconductor Fermi energy. The spectral spin current
into the metal is therefore
\begin{equation}
eI_{s}\left(  \varepsilon\right)  =G_{s}^{I}\left(  \varepsilon\right)
f_{s}\left(  \varepsilon\right)  ,
\end{equation}
where $G_{s}^{I}\left(  \varepsilon\right)  $ is the interface conductance at
energy $\varepsilon$. The total current of spin $s$ is given integrating over
energy
\begin{equation}
I_{s}=\int I_{s}\left(  \varepsilon\right)  d\varepsilon.
\end{equation}
We assume a charge current bias of $I_{C}=I_{\uparrow}+I_{\downarrow}$ and
introduce the spin current $I_{z}=I_{\uparrow}-I_{\downarrow}.$ We assume in
the following that energy relaxation is fast, such that the distribution
function at the interface is thermalized with non-equilibrium chemical
potentials $\mu_{s}$. At low temperatures, assuming local charge neutrality
$\mu_{\uparrow}+\mu_{\downarrow}=0$ and an interface conductance that does not
vary rapidly on the scale of $\mu_{s}$:
\begin{align}
eI_{z}  &  =\int_{0}^{\varepsilon_{F}+\mu_{\uparrow}}G_{\uparrow}^{I}\left(
\varepsilon\right)  d\varepsilon-\int_{0}^{\varepsilon_{F}+\mu_{\downarrow}%
}G_{\downarrow}^{I}\left(  \varepsilon\right)  d\varepsilon\\
&  =eI_{z}^{\left(  0\right)  }+\frac{\mu_{z}}{2}G^{I}\left(  \varepsilon
_{F}\right)  ,
\end{align}%
\begin{equation}
eI_{C}=\int_{0}^{\varepsilon_{F}}G^{I}\left(  \varepsilon\right)
d\varepsilon+\frac{\mu_{z}}{2}p\left(  \varepsilon_{F}\right)  G^{I}\left(
\varepsilon_{F}\right)  \label{IC}%
\end{equation}
where $G^{I}=G_{\uparrow}^{I}+G_{\downarrow}^{I}\ ,$ $p=\left(  G_{\uparrow
}^{I}-G_{\downarrow}^{I}\right)  /G^{I},$ $\mu_{z}=\mu_{\uparrow}%
-\mu_{\downarrow}$ is the spin accumulation at the interface and%
\begin{equation}
eI_{z}^{\left(  0\right)  }=\int_{0}^{\varepsilon_{F}}p\left(  \varepsilon
\right)  G^{I}\left(  \varepsilon\right)  d\varepsilon.
\end{equation}
In the degenerate limit, assuming that the conductivity is proportional to the
density, the magnetically active region of the semiconductor is determined by
the up-stream spin-diffusion length $\ell_{\mathrm{u}}=\ell_{\mathrm{sd}%
}\left(  \sqrt{1+X^{2}}-X\right)  $, where $X=3eI_{C}/\left(  8G_{0}%
^{SC}\varepsilon_{F}\right)  $ is a measure of the potential drop induced by
the current over the (zero-bias) spin diffusion length $\ell_{\mathrm{sd}}$ in
terms of the linear bulk conductance $G_{0}^{SC}=S\sigma^{SC}/\ell
_{\mathrm{sd}}$ of a semiconductor cube with area $S$ and thickness
$\ell_{\mathrm{sd}}$ \cite{Flatte}. In spite of the reduced spin-diffusion
length, the conductance of the spin-coherent region is increased compared to
the zero-bias limit:
\begin{equation}
G^{SC}=G_{0}^{SC}\frac{\ell_{\mathrm{u}}}{\ell_{\mathrm{sd}}}%
\end{equation}
(not $G_{0}^{SC}\ell_{\mathrm{sd}}/\ell_{\mathrm{u}}\ $as might be expected
naively). We then arrive at the effective circuit in Fig. 2, according to
which the spin current $I_{z}=I_{\uparrow}-I_{\downarrow}$ that flows from the
semiconductor bulk to the interface reads
\begin{equation}
eI_{z}=\frac{eI_{z}^{\left(  0\right)  }}{1+G^{I}/2G^{SC}} \label{Iz}%
\end{equation}
and the sign of $\mu_{z}=-eI_{z}/G^{SC}$ is opposite to that of $I_{z}$. A low
conductance $G^{SC}\rightarrow0$ suppresses the spin current \cite{Schmidt00},
but not the spin accumulation! By reversing $I_{C}$ and keeping in mind that
the interface conductance is in general much smaller and less bias dependent,
similar equations hold as well for reversed-bias Schottky barriers. As
mentioned above, most experiments on reverse bias junctions focus on the spin
current. The conductance mismatch problem is reflected in Eq. $\left(
\text{\ref{Iz}}\right)  $, where a small semiconductor conductance is seen to
suppress the spin current. In Refs. \cite{Schmidt00,Fert01} it was pointed out
that a significantly polarized spin current can only be achieved when the
reverse-bias Schottky barrier conductance is sufficiently small. However, in
this case the spin accumulation $\mu_{z}$ is suppressed, which explains why
Stephens \textit{et al}. \cite{Stephens} only detected spin accumulation with
a forward bias.

We now turn to the spin-accumulation in the presence of a variable in-plane
magnetic field, taking the magnetization of F to be parallel to the
$z-$direction and the magnetic field $B$ in the $y-$direction. The
magnetic-field induced non-collinearity of spin accumulation and magnetization
creates a spin transfer torque on the ferromagnet, thus opens new decay
channels \cite{Bauer04} proportional to the spin-mixing conductance
$G_{\uparrow\downarrow}^{I}$ at the Fermi energy \cite{Brataas00}. The Bloch
equation for the spin accumulation $\left\langle \mu\right\vert =\left(
\mu_{x},\mu_{y},\mu_{z}\right)  \ $can be written
\begin{equation}
-T^{I}\frac{d|\mu\rangle}{dt}=\mathbf{\Gamma}|\mu\rangle+\frac{2e|I_{z}%
\rangle}{G_{I}}, \label{Kinetic}%
\end{equation}
where $T^{I}=2e^{2}\mathcal{D}/G^{I}$ is the interface relaxation time in
terms of the (single spin) semiconductor energy density of states
$\mathcal{D}$ in the magnetically active volume.%
\begin{equation}
\mathbf{\Gamma}=\left(
\begin{array}
[c]{ccc}%
\eta_{r}+\xi & \eta_{i} & T^{I}\omega\\
-\eta_{i} & \eta_{r}+\xi & 0\\
-T^{I}\omega & 0 & 1+\xi
\end{array}
\right)
\end{equation}
where $\eta_{r}=2\operatorname{Re}G_{\uparrow\downarrow}^{I}/G^{I},$ $\eta
_{i}=2\operatorname{Im}G_{\uparrow\downarrow}^{I}/G^{I},$ $\xi=2G^{SC}%
/G^{I}\ $and the Larmor frequency $\omega=g_{e}\mu_{B}B/\hbar\ $in terms of
the g-factor $g_{e}$ and the Bohr magneton $\mu_{B}.$ Eq. (\ref{Kinetic})
holds when the relaxation rate of the electron orbital degrees of freedom is
sufficiently larger than $\omega$. The source term is the current bias applied
to the semiconductor. $\left\langle I_{z}\right\vert =\left(  0,0,I_{z}%
^{\left(  0\right)  }\right)  .$ The stationary state solution for the Bloch
equation, $|\mu\rangle=\mathbf{\Gamma}^{-1}2e|I\rangle/G_{I},$ is easily
obtained analytically. The spin accumulation at the interface reads:
\begin{equation}
\left\langle \mu\right\vert =\frac{\left(
\begin{array}
[c]{ccc}%
-\left(  \eta_{r}+\xi\right)  \omega/T^{I}, & \eta_{i}\omega/T^{I}, & -\left(
\eta_{r}+\xi\right)  ^{2}-\eta_{i}^{2}%
\end{array}
\right)  }{\left[  \left(  \eta_{r}+\xi\right)  ^{2}+\eta_{i}^{2}\right]
\left(  1+\xi\right)  +\left(  \eta_{r}+\xi\right)  \omega^{2}}\frac
{2eI_{z}^{\left(  0\right)  }}{G^{I}}%
\end{equation}
Stephens \textit{et al}. \cite{Stephens} found the component of the spin
accumulation normal to the interface $\mu_{x}$ well represented by a
Lorentzian $A\omega/\left(  \omega^{2}+T^{-2}\right)  $. This form also
follows from our rate equations with%
\begin{equation}
\left(  \frac{T^{I}}{T}\right)  ^{2}=\left[  \left(  \eta_{r}+\xi\right)
^{2}+\eta_{i}^{2}\right]  \frac{1+\xi}{\eta_{r}+\xi}%
\end{equation}
and $A=-2eI_{z}^{\left(  0\right)  }/\left(  G^{I}T^{I}\right)  .$ In the
limit of a highly resistive semiconductor, $\xi\ll1$, and taking $\eta
_{i}=0,\eta_{r}=1$, we find that $T\rightarrow T^{I}$ and $AT\rightarrow
\mu_{z}\left(  \omega=0\right)  ,$ \textit{i.e.} the zero field spin
accumulation. It is therefore possible to obtain information about the
interface conductance from the experimental spin dephasing time.

The MnAs%
$\vert$%
GaAs systems has been studied intensively \cite{Tanaka}, but not much is known
about the electronic transport properties. Epstein \textit{et al}.
\cite{Epstein1} reported that the conductance polarization is opposite to the
magnetization direction, \textit{i.e}. $p<0$. We compute MnAs%
$\vert$%
GaAs (100) interface conductances $G^{I}\left(  \varepsilon\right)  $ for the
ZincBlende $\left(  \alpha\right)  $ structure\cite{beta} by scattering matrix
calculations with a first-principles tight-binding basis \cite{Zwierzycki},
assuming flat-band conditions with a Schottky barrier height of 0.8 eV (see
Fig. 3). We find large differences between clean, and on a monolayer scale,
alloy-disordered interfaces, \textit{e.g}. the interface polarization changes
sign when the interface becomes increasingly dirty. The negative polarization
found in \cite{Epstein1} is thus consistent with non-ideal interfaces. These
features are quite similar to results for Fe%
$\vert$%
InAs that does not have a Schottky barrier \cite{Zwierzycki}. We also note
that in the regime considered here we calculate an $\eta_{r}\simeq1$ in all
cases and $\eta_{i}\simeq0\ (0.35)$ for clean (disordered) interfaces. We
parametrize the estimated interface conductance in terms of the SC Sharvin
conductance$\ G_{Sh}\left(  \varepsilon_{F}\right)  /S=\left(  2e^{2}%
/h\right)  \left(  2m^{\ast}\varepsilon_{F}\right)  /\left(  4\pi\hbar
^{2}\right)  $ times a transparency parameter that at 12 meV is found to be
$\varkappa=0.27$ for clean and dirty interfaces$.$

In order to make contact with Stephens \textit{et al}. \cite{Stephens} the
above results for the planar junction have to be adapted to the experimental
geometry in Fig. 1. For GaAs with doping density $n=10^{17}%
\operatorname{cm}%
^{-3},$ we take a mobility 3000 cm$^{2}/\left(  \mathrm{Vs}\right)  $, an
effective mass $m^{\ast}=0.067m_{e},$ and spin-flip diffusion length
$\ell_{\mathrm{sd}}\simeq2%
\operatorname{\mu m}%
$ \cite{Kikkawa}, that is significantly larger than the film thickness
$d^{SC}=0.5%
\operatorname{\mu m}%
$. The measured excess spin-dephasing rates $T^{-1}$ at applied currents
$I_{C}\ $are listed in the table. Close to the interface, the up-stream
spin-flip diffusion length is not significantly reduced, so not only the whole
$\left(  \simeq5\times50%
\operatorname{\mu m}%
^{2}\right)  $ area under the conducting contact is spin coherent at all
currents, but also strips on both sides with widths of the order of
$\ell_{\mathrm{sd}}$ and $\ell_{\mathrm{u}}$, respectively. The drift effect
on the available density of states is small, but it significantly affects the
resistance of the spin coherent region. Due to the thin layer thickness of the
GaAs, lateral spin diffusion may be disregarded. The results in the table are
obtained assuming that $\eta_{i}=0,\eta_{r}=1.$

\begin{table}[ptb]
\begin{center}%
\begin{tabular}
[c]{lllll}%
$I_{C}\left(  \operatorname{mA}\right)  $ & $\frac{1}{T}\left(
\operatorname{ns}^{-1}\right)  $ & $R^{SC}\left(  \Omega\right)  $ &
$SR^{I}\left(  \frac{\text{f}\Omega\operatorname{m}^{2}}{10^{-5}}\right)  $ &
$\kappa$\\
$0.3$ & $0.25$ & $43$ & $25$ & $0.004$\\
$1.1$ & $1.2$ & $63$ & $5.2$ & $0.014$\\
$2.7$ & $6$ & $111$ & $1.1$ & $0.074$%
\end{tabular}
\end{center}
\caption{Experimental results for the spin dephasing rate $1/T$ at selected
current bias $I_{C}$ from Ref. \onlinecite{Stephens} and the estimates of
device parameters according to the discussion in the text. $R^{SC}$ and
$SR^{I}$ are the $SC$\ bulk resistance of the magnetically active region and
the interface resistance for the given bias. $\kappa$ is the transparency
parameter that measures the interface conductance in unit of the $SC$ Sharvin
conductance.}%
\end{table}

The experimental longevity of spins is striking and can only be explained by a
reduced interface conductance. Any spin-flip process disregarded here, caused,
\textit{e.g}., by heating \cite{Kikkawa} due to high currents, would
correspond to even smaller transparencies. The small $\kappa$ at small bias
reflects the residual Schottky barrier that is not yet completely screened. At
higher bias these remnants should disappear and the spin-dephasing time should
be governed by the intrinsic interface. At higher bias the interface
conductances deduced from the experiments grow to about one third of the
intrinsic first-principles results. An energy-averaged interface transparency
is accessed by the electrical current itself. Disregarding the small term
proportional to $\mu_{z}$ in Eq. (\ref{IC}), the current according to the
first-principles conductances and a contact area of $250%
\operatorname{\mu m}%
^{2}$ should be $I_{C}\simeq25%
\operatorname{mA}%
$. At the experimental currents of $I_{C}\lesssim3%
\operatorname{mA}%
$ the \emph{average} $\kappa$ is thus smaller than that at the Fermi energy
obtained from the Hanle effect. This can be explained by an energy dependent
$\kappa$ that decreases strongly when approaching the band edge. These
remaining puzzles might be related to the measured \cite{Stephens} spatial
inhomogeneity of the current induced spin accumulation and thus interface conductance.

Stephens \textit{et al}. \cite{Stephens} estimate the spin accumulation to be
10\% \ of\ the Fermi energy from nuclear polarization data compared to an
estimate of $\sim$15\% based on the first-principles results for disordered
interfaces. The spin accumulation is found to saturate and even decrease again
with large $I_{C}.$ A probable reason is a reduced spin-flip diffusion length,
either by heating or by a large drift contribution at higher bias.

In summary, we demonstrate how a transport property, the semiconductor%
$\vert$%
ferromagnet interface conductance, can be measured optically on an absolute
scale. The conductance mismatch is found to favor spin injection into
semiconductors in forward-biased magnetic Schottky barriers. At low bias the
interface-mediated spin-decay is much weaker in the experiments
\cite{Stephens} than expected from intrinsic SC%
$\vert$%
F interfaces. At higher bias the agreement becomes better, indicating that the
interface approaches (but does not reach) the Ohmic limit as calculated from
first-principles. Experiments that determine the spin accumulation on an
absolute scale would be of great help to refine the present analysis. The
observed negative polarization \cite{Epstein1,Stephens} can be explained by
disorder at the interfaces. A systematic study as a function of semiconductor
thickness could shed more light on the spin-decoherence in the non-linear
transport regime.

We thank Georg Schmidt, David Awschalom, and Jason Stephens for helpful
discussions. This work has been supported by the FOM Foundation and the EU
Commission FP6 NMP-3 project 505587-1 \textquotedblleft
SFINX\textquotedblright.

\begin{figure}[ptb]
\includegraphics[scale=0.6,trim=0 100 0 0]{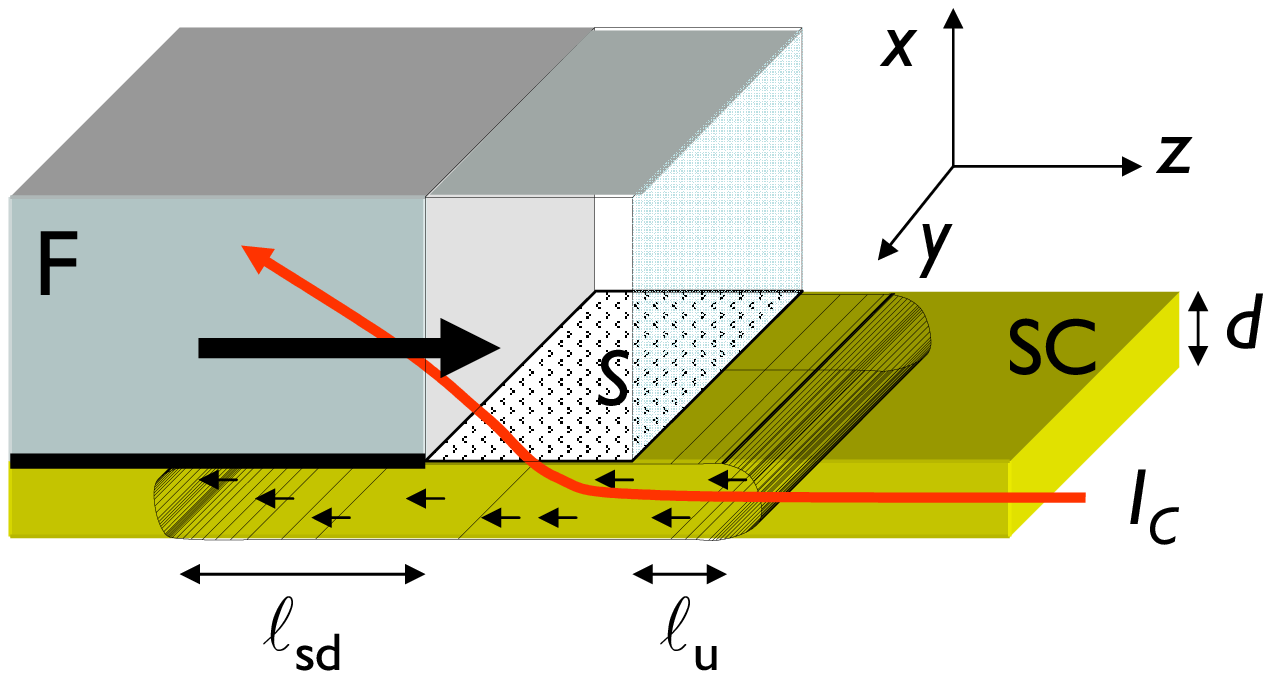}\caption{Schematic
drawing of the magnetic forward-biased Schottky diode of Stephens \textit{et
al}. \cite{Stephens}. The particle current $I_{C}$ is injected from the
semiconductor film SC of thickness $d$ into the ferromagnet F through a
contact area $S$. The excited spin accumulation diffuses back into the
semiconductor over the spin-diffusion length $\ell_{\mathrm{sd}}$ without bias
and up-stream diffusion length $\ell_{\mathrm{u}}$ against the bias. The spin
accumulation in the semiconductor is plotted for $p>0.$}%
\end{figure}

\begin{figure}[ptb]
\includegraphics[scale=0.6,trim=0 100 0 0]{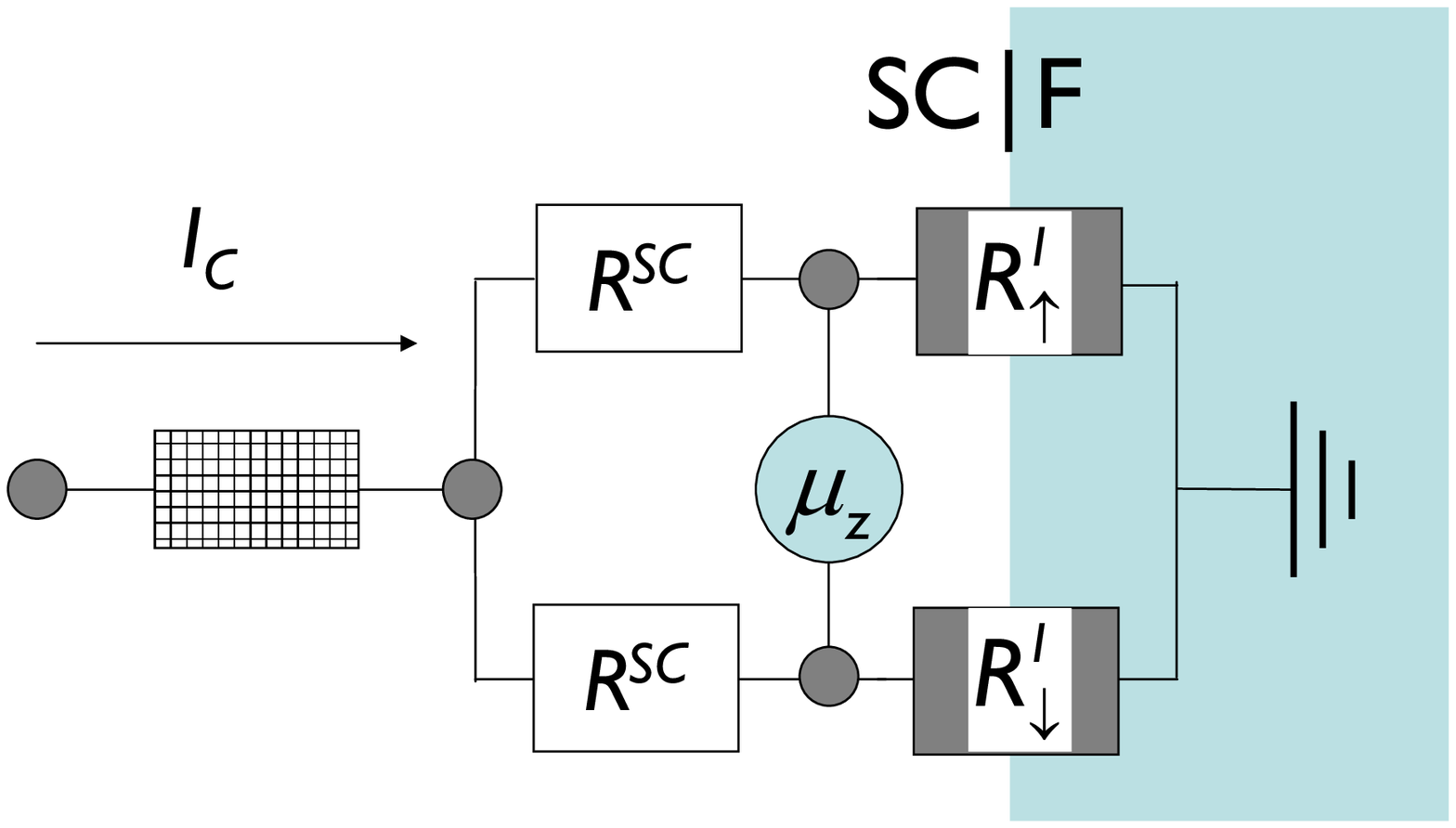}\caption{Magnetoelectronic
circuit for a current biased magnetic Schottky barrier in the absence of a
magnetic field. }%
\end{figure}

\begin{figure}[ptb]
\includegraphics[scale=0.6,trim=0 100 0 0]{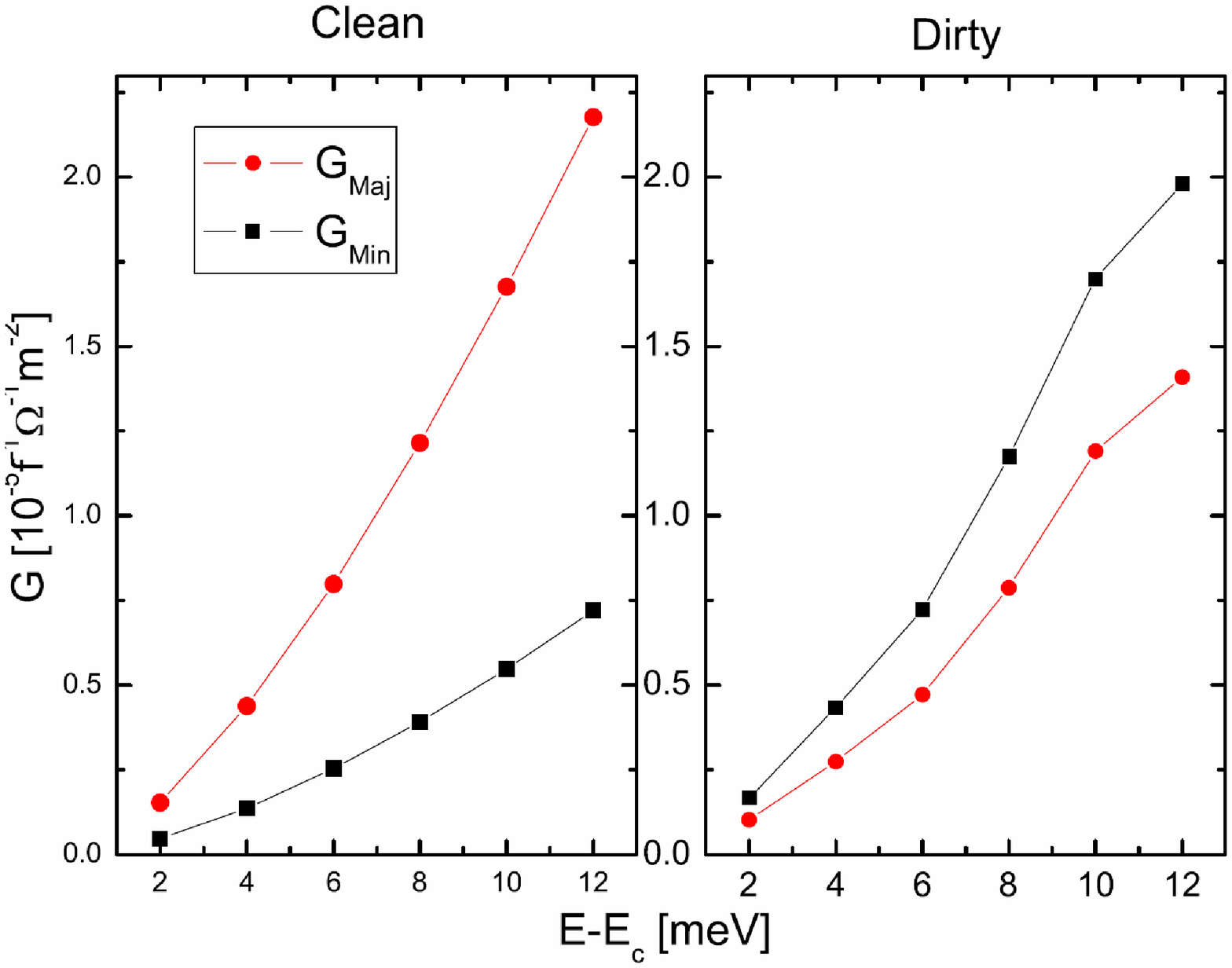}\caption{Intrinsic
conductance of a specular (left) and disordered (right) $\alpha-$MnAs/GaAs
(100) interface at flat band conditions as a function of the excess energy in
the GaAs conductance band and calculated from first principles
\cite{Zwierzycki}.}%
\end{figure}


\begin{thebibliography}{99}                                                                                               %


\bibitem {Schmidt00}G. Schmidt, D. Ferrand, and L. W. Molenkamp, A. T. Filip,
and B. J. van Wees, Phys. Rev. B \textbf{62}, 4790 (2000).

\bibitem {Wunnicke}O. Wunnicke, Ph. Mavropoulos, R. Zeller, P.H.Dederichs, and
D. Grundler, Phys. Rev. B \textbf{65}, 241306 (2002).

\bibitem {Zwierzycki}M. Zwierzycki, K. Xia, P. J. Kelly, G.E.W. Bauer, and I.
Turek, Phys. Rev. B \textbf{67}, 092401 (2003).

\bibitem {Fiederling99}R. Fiederling, M. Keim, G. Reuscher, W. Ossau, G.
Schmidt, A. Waag, and L. W. Molenkamp, Nature \textbf{402}, 787 (1999); Y.
Ohno, D. K. Young, B. Beschoten, F. Matsukura, H. Ohno, and D. D. Awschalom,
Nature \textbf{402}, 790 (1999).

\bibitem {Ohno}H. Ohno, Institute of Physics Conference Series \textbf{171},
37 (2003).

\bibitem {Brataas00}A. Brataas, Yu.V. Nazarov, and G.E.W. Bauer, Phys. Rev.
Lett. \textbf{84}, 2481 (2000); Eur. Phys. J. B \textbf{22}, 99 (2001).

\bibitem {Brataasb}A. Brataas, Y. Tserkovnyak, G.E.W. Bauer, and B.I.
Halperin, Phys. Rev. B \textbf{66}, 060404 (2002).

\bibitem {Rashba00}E.~I.~Rashba, Phys.~Rev.~B \textbf{62}, R16267 (2000).

\bibitem {Fert01}A.~Fert and H.~Jaffres, Phys.~Rev.~B \textbf{64}, 184420 (2001).

\bibitem {Monsma}D.J. Monsma, J.C. Lodder, Th.J.A. Popma, and B. Dieny, Phys.
Rev. Lett. \textbf{74}, 5260 (1995); O. M. J. van't Erve, R. Vlutters, P. S.
Anil Kumar, S. D. Kim, F. M.Postma, R. Jansen, and J. C. Lodder, Appl. Phys.
Lett. \textbf{80}, 3787 (2002).

\bibitem {Parkin}S. van Dijken, X. Jiang. and S.P. Parkin, Appl. Phys. Lett.
\textbf{83}, 951 (2003); I. Appelbaum, K. J. Russell, D. J. Monsma, V.
Narayanamurti, C. M. Marcus, M. P. Hanson, and A. C. Gossard; Appl. Phys.
Lett. 83, 4571 (2003)

\bibitem {Jonker}O.M.J. van 't Erve, G. Kioseoglu, A.T. Hanbicki, C.H. Li, and
B.T. Jonker, R. Mallory, M. Yasar, and A. Petrou, Appl. Phys. Lett.
\textbf{84}, 4334 (2004) and references therein.

\bibitem {Stephens}J. Stephens, J. Berezovsky, J. P. McGuire, L. J. Sham, A.
C. Gossard, and D. D. Awschalom, Phys. Rev. Lett \textbf{93}, 097602 (2004).

\bibitem {Bauer04}G.E.W. Bauer, A. Brataas, Y. Tserkovnyak, B. I. Halperin, M.
Zwierzycki, and P. J. Kelly, Phys. Rev. Lett. 92, 126601 (2004); M. Zaffalon
and B.J. van Wees, cond-mat/0411407.

\bibitem {Tanaka}M. Tanaka, Sem. Sc. Techn. \textbf{17}, 327 (2002).

\bibitem {Kikkawa}J. M. Kikkawa and D. D. Awschalom, Phys. Rev. Lett.
\textbf{80}, 4313 (1998).

\bibitem {Flatte}Z. G. Yu and M. E. Flatt\'{e}, Phys. Rev. B \textbf{66},
235302 (2002).

\bibitem {Epstein1}R.J. Epstein, I. Malajovich, R.K. Kawakami, Y. Chye, M.
Hanson, P.M. Petroff, A.C. Gossard, and D.D. Awschalom, Phys. Rev. B
\textbf{65}, 121202 (2002).

\bibitem {beta}Preliminary results on geometry optimized $\beta$-MnAs%
$\vert$%
GaAs interface show interesting differences, such as a negative polarization
for specular interfaces and a stronger dependence of $\kappa$ on disorder
morphology \ (P.-X. Xu, J. Ren, K. Xia \textit{et al.}, unpublished).
\end{thebibliography}
\end{document}